\documentclass[journal,twocolumn]{IEEEtran}
\usepackage{epsfig,makeidx,color}
\usepackage{amsmath,amssymb,amsthm,bbm}
\usepackage{cite,graphicx}
\usepackage{enumerate}
\usepackage{hyperref}
\hypersetup{
	colorlinks = true,
	citecolor=blue,
}

\def\defh{
\mbox{\footnotesize $ \begin{array}{c} H_1 \cr > \cr < \cr H_0 \end{array} $}}

\def\cH{{\cal H}}
\def\cK{{\cal K}}

\def\cG{{\cal G}}
\def\cQ{{\cal Q}}

\def\rH{{\rm H}}

\def\rT{{\rm T}}

\def\uR{{\mathbb R}}
\def\uC{{\mathbb C}}
\def\uE{{\mathbb E}}

\DeclareMathOperator*{\argmin}{\arg\!\min}
\DeclareMathOperator*{\argmax}{\arg\!\max}
\def\indicator{\mathbbm{1}}

\def\be{ \begin{equation} }
\def\ee{ \end{equation} }
\def\bea{ \begin{eqnarray} }
\def\eea{ \end{eqnarray} }
\def\bx{{\bf x}}
\def\by{{\bf y}}
\def\bc{{\bf c}}

\def\bb{{\bf b}}

\def\bs{{\bf s}}
\def\ba{{\bf a}}
\def\br{{\bf r}}

\def\bm{{\bf m}}
\def\bn{{\bf n}}

\def\bp{{\bf p}}

\def\bv{{\bf v}}
\def\bw{{\bf w}}

\def\bB{{\bf B}}
\def\bC{{\bf C}}

\def\bI{{\bf I}}

\def\bR{{\bf R}}

\def\b0{{\bf 0}}

\def\bPsi{{\bf \Psi}}

\def\cC{{\cal C}}
\def\cD{{\cal D}}
\def\cK{{\cal K}}

\def\cI{{\cal I}}
\def\cN{{\cal N}}

\ifCLASSOPTIONonecolumn
  \newcommand{\figwidth}{0.60\columnwidth}
\else
  \newcommand{\figwidth}{0.87\columnwidth}
\fi

\begin{document}

\title{A Cross-Layer Approach to
Data-aided Sensing using Compressive Random Access}

\author{Jinho Choi\\
\thanks{The author is with
the School of Information Technology,
Deakin University, Geelong, VIC 3220, Australia
(e-mail: jinho.choi@deakin.edu.au).}}

\date{today}
\maketitle

\begin{abstract}
In this paper, data-aided sensing
as a cross-layer approach in Internet-of-Things (IoT) applications
is studied,
where multiple IoT nodes collect  measurements and transmit
them to an Access Point (AP).
It is assumed that 
measurements have a sparse representation (due to spatial correlation)
and the notion of Compressive Sensing (CS) can be exploited
for efficient data collection. For data-aided sensing,
a node selection criterion is proposed to efficiently
reconstruct a target signal through iterations
with a small number of measurements from selected nodes.
Together with Compressive Random Access (CRA) 
to collect measurements from nodes, 
compressive transmission request is proposed to efficiently
send a request signal to a group of selected nodes.
Error analysis on compressive transmission request 
is carried out and the impact of errors
on the performance of data-aided sensing is studied.
Simulation results show that data-aided sensing
allows to reconstruct the target information 
with a small number of active nodes and is robust to
nodes' decision errors on compressive transmission request.
\end{abstract}

{\IEEEkeywords
Data-aided sensing;
Internet-of-Things (IoT); Cross-Layer}

\ifCLASSOPTIONonecolumn
\baselineskip 26pt
\fi

\section{Introduction} \label{S:Intro}

There have been extensive efforts to realize the idea
of the Internet of Things (IoT) 
as the IoT has a huge impact on a number of applications
ranging from environmental monitoring
to factory automation to smart cities
and fosters new business growth
\cite{Atzori10} \cite{Gubbi13} \cite{Fuqaha15} \cite{Kim16}.
By providing the connectivity for devices (e.g., sensors
and actuators), a number of IoT applications can be
emerged.
For example, in \cite{Kelly13},
an IoT-based implementation for 
environmental condition monitoring in homes
is studied.

To implement IoT systems, various reference
architectures for IoT infrastructure are proposed
\cite{Fuqaha15}. In general,
the bottom layer, called objects or perception layer, 
is to collect and process information or data from devices.
This is often divided into two sub-layers, namely
connectivity layer and data layer, which 
are responsible for
collecting data and processing collected data, respectively,
in an IoT system.
The two layers
are usually decoupled for ease of implementation as in \cite{Kuo18}, 
where an IoT platform is used as a Wireless
Sensor Network (WSN) for environmental sensing
and data delivering.

\begin{figure}[thb]
\begin{center}
\includegraphics[width=\figwidth]{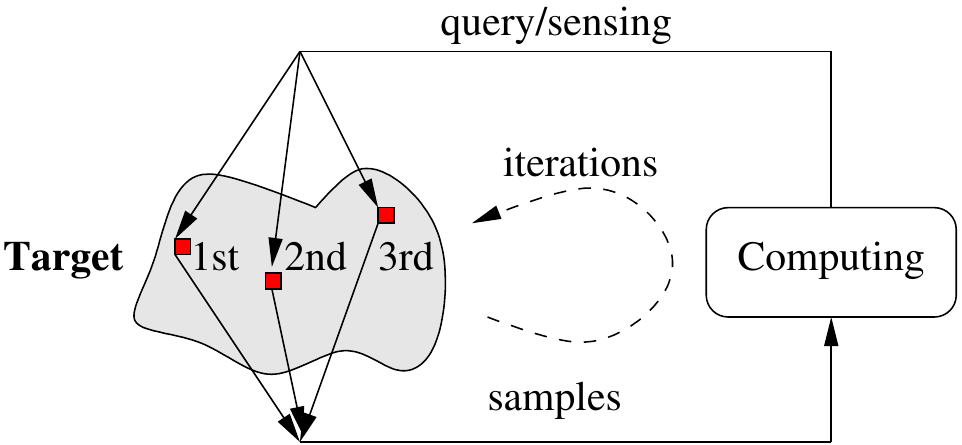}
\end{center}
\caption{IoT sensing and computing to acquire information,
which can be iteratively performed for better acquisition/sensing.}
        \label{Fig:das}
\end{figure}

In this paper, however, a cross-layer
design to combine data sensing and processing
is considered to make data-aided sensing possible.
Data-aided sensing is a novel approach
that can actively choose IoT devices/sensors
with desired information
or measurements by analyzing existing data sets
to improve the overall quality of data sets in iterations.
For example, as shown in Fig.~\ref{Fig:das}, 
the next sensing point or sensor can be decided by analyzing
the existing data set 
for a better outcome provided that the system allows
to choose sampling points or IoT devices/sensors at certain locations,
while the first sampling point can be randomly decided.
Through iterations in data-aided sensing,
the outcome can be improved.
As a result, the number of iterations with data-aided
sensing can be smaller than that without data-aided sensing
to reach a certain desired quality of data set.

Certainly, cross-layer design for data-aided sensing 
may not be suitable to every IoT systems. 
In general, for efficient data-aided
sensing, certain features of connectivity layer 
are to be supported.
It is expected that devices are able to directly connect to 
a Base Station (BS) or Access Point (AP),
which requires long-range connectivity with star topology.
A number of approaches are considered to support long-range
connectivity such as
Low-Power Wide Area Networking (LP-WAN) and 
Machine-Type Communication (MTC) \cite{Sanch16} \cite{Bockelmann16}.

In MTC, in order to support 
a large number of devices with sparse activity, 
the notion of Compressive Sensing (CS) \cite{Donoho06} \cite{Candes06}
\cite{Eldar12}
has been adopted,
which results in
Compressive Random Access (CRA) \cite{Wunder14} \cite{Choi17} 
\cite{Choi_CRA18} \cite{Choi18}.
In CRA, multiple IoT devices/sensors
with unique signature sequences 
can simultaneously transmit, and 
a receiver uses
a CS-based Multi-User Detection (MUD) algorithm
to detect their activity as well as their 
signals \cite{Zhu11} \cite{Applebaum12}.
In this paper, CRA is assumed for connectivity layer
as it can also be used for active data collection
by exploiting the assignment of a unique signature sequence to each device
in data-aided sensing.

For data-aided sensing, in this paper,
a specific application where the AP aims to reconstruct 
a target signal from as few devices' measurements as possible 
is mainly considered under the assumption that the 
set of measurements has a sparse representation.
In WSNs, when the measurements of sensors are spatially correlated,
they can have a sparse representation
and the notion of CS can be exploited for efficient data collection
\cite{Leinonen15} \cite{Masoum17} \cite{Wu18}.
In this case, 
thanks to data-aided sensing, 
measurements from only a fraction of devices may be needed 
for the AP to have a full picture
(a reconstructed target signal), which may result in
a high
energy efficiency as well as a short reconstruction time.

The rest of the paper is organized as follows.
In Section~\ref{S:SM}, a signal model
with a sparse representation 
is presented to study data-aided sensing.
The reconstruction of
a target signals from randomly collected
measurements (based on the notion
of CS) and CRA with multiple nodes 
are discussed in Section~\ref{S:RS}.
In Section~\ref{S:DAS}, two different approaches
to collect measurements are studied and the node selection
criterion for data-aided sensing is proposed.
To collect measurements from the nodes 
that are chosen according to the node selection criterion,
a simple but efficient approach 
for controlled access is developed in Section~\ref{S:CA}
and its error analysis is presented in Section~\ref{S:EA}.
Simulation results are shown in Section~\ref{S:Sim}
and the paper is concluded with remarks
in Section~\ref{S:Conc}.

{\it Notation}:
Matrices and vectors are denoted by upper- and lower-case
boldface letters, respectively.
The superscripts $\rT$ and $\rH$
denote the transpose and complex conjugate, respectively.
The $p$-norm of a vector $\ba$ is denoted by $|| \ba ||_p$
(If $p = 2$, the norm is denoted by $||\ba||$ without the subscript).
The support of a vector is denoted by ${\rm supp} (\bx)$.
$\uE[\cdot]$
and ${\rm Var}(\cdot)$
denote the statistical expectation and variance, respectively.
$\cC\cN(\ba, \bR)$
represents the distribution of
Circularly Symmetric Complex Gaussian (CSCG)
random vectors with mean vector $\ba$ and
covariance matrix $\bR$.
The Q-function is given by
$\cQ(x) = \int_x^\infty \frac{1}{\sqrt{2 \pi} } e^{- \frac{t^2}{2} } dt$.

\section{Signal Model for IoT Data Collection} \label{S:SM}

Suppose that a number of IoT nodes\footnote{Throughout
the paper, it is assumed 
that nodes, devices, and sensors are interchangeable.}
are deployed in a 
certain site or cell 
to collect and transmit measurements to
an AP
for IoT sensing (or data collection) in a certain 
application (e.g., for environmental monitoring).
Denote by $K$ the number of IoT nodes and by
$\bm_k$ the measurement at node $k
\in \cK = \{ 1, \ldots, K\}$.
The AP can collect measurements from the nodes 
and process them to create a target information for the site.
For example, if each node sends the temperature at each
location, the AP can build a heat map for the site.

In this paper,
it is assumed that
the signal vector obtained
from all the nodes' measurements has a sparse representation 
\cite{Frag13} \cite{Karakus13} \cite{Wu18}.
Let
$\cD = \{\bm_k, k \in \cK\}$.
For the sake of simplicity,
the target signal is assumed to be a 
$K \times 1$ vector, denoted by $\bv$, 
as follows:
$$
[\bv]_k = m_k \in \uR, \ k \in \cK.
$$
Furthermore, $\bv$ can be represented by 
a sparse vector $\bs \in \uR^{M \times 1}$. 
To this end, it is assumed that
\be
v_k = \bb_k^\rT \bs,  \ k \in \cK,
	\label{EQ:vbs}
\ee
where $\bb_k \in \uR^{M \times 1}$ is the measurement
vector at node $k$, which is known at the AP.

From \eqref{EQ:vbs}, $\bv$ becomes
$\bv = \bB^\rT \bs$,
where $\bB = [\bb_1 \ \ldots \ \bb_K]^\rT \in \uR^{M \times K}$.
Throughout the paper, it is assumed that
$||\bs||_0 = S$,
i.e., $\bs$ is a $S$-sparse vector, where $S \ll K$.
As a result, it is possible to obtain $\bv$ once 
$\bs$ is known.
In other words, it might be possible to estimate
$\bv$ without knowing all the nodes' measurements.

\begin{figure}[thb]
\begin{center}
\includegraphics[width=\figwidth]{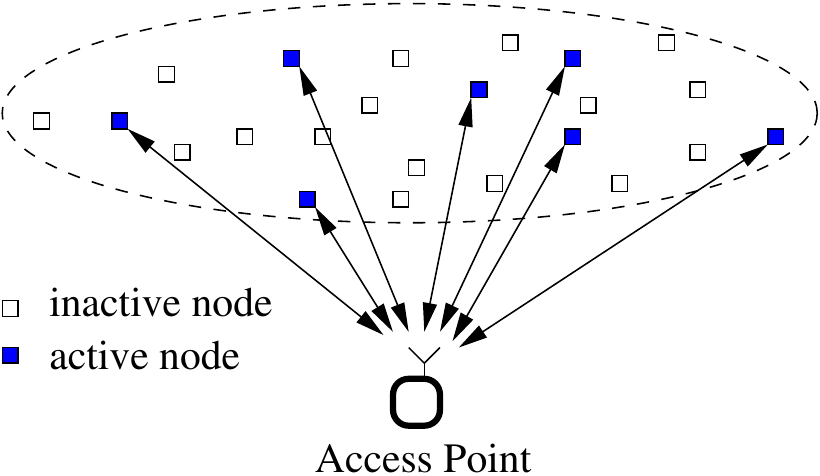}
\end{center}
\caption{A site with a number of nodes or sensors 
deployed to perform and transmit measurements
to an AP. Each node has 
its measurement $\bm_k$ and can be
characterized by its measurement vector, $\bb_k$, which is known at the AP.
In addition, it is assumed that bi-directional communication is possible.}
        \label{Fig:sys}
\end{figure}

For data-aided sensing, with multiple rounds of sensing,
multiple access is employed
to receive multiple measurements
in each round as in Fig.~\ref{Fig:sys}.
Let $\cD_0$ denote the initial set of measurements associated with 
a subset of $\cK$, denoted by $\cI_0$.
For convenience, let $|\cI_0| = N$, where $N \ll K$, and
denote by $\cI_q$ the index set of active nodes in round $q$
throughout the paper.
Since no data sets are available in round 0, $\cI_0$ can be a random
index set. Once $\cD_0$ is available at the AP from random
active nodes as in Fig.~\ref{Fig:sys},
the AP can determine the next index set of active nodes that can 
improve the quality of data set
(i.e., the estimation accuracy of $\bv$), which is denoted by
$\cI_1$. Through iterations, 
the AP can have a (growing) data set
that can effectively improve the estimate of $\bv$.

\section{Random Sensing via Compressive Random Access}	\label{S:RS}

Since $S \ll K$, in order to find $\bs$,
it may not be necessary to ask all the nodes to transmit
their measurements.
That is, provided that the support of $\bs$ is known,
the AP only needs to have measurements from $S$ nodes,
In this case, the time to acquire $\bv$ becomes shorter 
by a factor of $\frac{K}{S}$.
In addition, the life time of all IoT nodes increases.
However, the support of $\bs$ is unknown and 
more than $S$ nodes need to  transmit
their  measurements. According to 
\cite{Donoho06} \cite{Candes06} \cite{Eldar12},
it might be still possible to keep the number of the nodes
transmitting measurements small by exploiting the sparsity
of $\bs$ to reconstruct $\bv$ based on 
the notion of CS. In this section, 
random sensing or projection via random access 
is studied for the initial sensing.

\subsection{Random Sensing}

Suppose that each node can decide whether or not it transmits
its measurement randomly in the initial round, 
i.e., round 0. In this case,
as shown in Fig.~\ref{Fig:sys}, a fraction of nodes become
active (at a time) and transmit their measurements.
Then, the AP can have
a random subset of $\bv$ and the resulting operation
is referred to as random sensing
in this paper.

Denote by $\cI = \cI_0$ the set of the
active devices transmitting their signals,
which is a  random index set of $N \ (\le K)$ active nodes
because the AP does not know which nodes are active or not.
To be precise, let
\be
\cI = \{i_1, \ldots, i_N\}, \ i_n \in \cK,
\ee
where $i_n$ represents the index of the $n$th active node
(thus, the $i_n$'s are unique).
The subvector of $\bv$ associated with $\cI$ is denoted by
$\bw$
and let $\bPsi = [\bb_{i_1} \ \cdots \ \bb_{i_N}]^\rT$. 
Then, it can be shown that
\be
\bw = \bPsi \bs \in \uR^{N \times 1},
	\label{EQ:wPs}
\ee
which can be seen as a random sensing or sampling of $\bv$.
It is known that if 
$N \ge C S \log \left( \frac{M}{S} \right)$,
where $C$ is a constant, $\bs$ can be recovered
from $\bw$ under certain conditions of $\bPsi$ in \eqref{EQ:wPs}
\cite{Donoho06} \cite{Candes06} \cite{Eldar12}.
Once $\bs$ is available, 
$\bv$ can be obtained from \eqref{EQ:vbs}.
In other words, without collecting all the measurements
from $K$ nodes, it is possible to estimate
$\bv$ from $N$ nodes' measurements (i.e., $\bw$).

\subsection{Compressive Random Access}	\label{SS:CRA}

For random sensing, each node can locally
decide whether or not to transmit with a certain access probability,
denoted by $p_{\rm a}$,
which results in random access.
In this subsection, a random access method 
is discussed with a unique signature for each node 
so that the AP knows the node associated with each measurement 
when it recovers measurements from the received signal.

Suppose that each node has a unique signature sequence
consisting $L$ elements,
denoted by $\bc_k \in \uC^{L \times 1}$.
For normalization purposes, 
it is assumed that $||\bc_k ||^2 = 1$ for all
$k$.
In addition,
node $k$ transmits its signal
to deliver the measurement $\bm_k$ to
the AP. Then, the received 
signal at the AP becomes
\be
\br_t = \sum_{k \in \cI} h_k \sqrt{P_k} \bc_k d_{k,t} + \bn_t,
\ t = 0, \ldots, T-1,
	\label{EQ:br}
\ee
where $\bn_t \sim \cC \cN(\b0, N_0 \bI)$ is the background noise
sequence and $h_k$ represents the channel coefficient
from node $k$ to the AP, and $P_k$ represents the transmit power.
Here, $d_{k,t}$ represents the $t$th data symbol and $T$ denotes
the length of packet.
The resulting system can be seen as a code
division multiple access \cite{VerduBook} \cite{ChoiJBook}
system since $\bc_k$ is not only a signature
sequence, but also a spreading sequence as shown in \eqref{EQ:br}.
Thus, it is possible 
for the AP 
to decode the signals when multiple nodes transmit their measurements
simultaneously.

From \eqref{EQ:br}, the received signal at the AP is re-written as
\begin{align}
\br_t = \bC \bx_t + \bn_t,
	\label{EQ:brC}
\end{align}
where $\bC = [\bc_1 \ \ldots \ \bc_K]$,
$\bx_t = [x_{1,t} \ \ldots \ x_{K,t}]$, and
$x_{k,t} = h_k \sqrt{P_k} a_k d_{k,t}$.
Here, $a_k \in \{0,1\}$ is the activity variable for node $k$.
Throughout the paper, Time Division Duplexing (TDD)
is assumed
so that the channel reciprocity holds.
Thus, when the AP transmits a beacon signal, it can 
also be used as a pilot signal that allows each node to estimate
$h_k$. With known channel gain,
the transmit power at node $k$ can be decided as 
\be
P_k = \frac{P_{\rm rx}}{|h_k|^2},
	\label{EQ:Pk}
\ee
where $P_{\rm rx}$ is the required received signal
power at the AP. In \eqref{EQ:brC}, 
$\cI = {\rm supp}(\ba)$,
where $\ba = [a_1 \ \ldots \ a_K]^\rT$.
Since the transmit power at each node is limited,
if $P_k > P_{\rm max}$
or 
\be
|h_k|^2 < \frac{P_{\rm rx}}{P_{\rm max}} = \omega, 
	\label{EQ:homega}
\ee
where $P_{\rm max}$ is the maximum
transmit power, node $k$ cannot transmit its measurement.
In this case, it is assumed that
$a_k$ becomes 0 (as node $k$ is not active) and
the resulting power control policy
becomes the truncated channel inversion power control 
policy \cite{Goldsmith97a}.

Since $h_k$ is random and some nodes may not be active
when in sleep mode,
$a_k$ becomes an independent Bernoulli random variable. 
Thus, in this paper, it is assumed that
$\Pr(a_k = 1) = p_{\rm a}$
and $\Pr(a_k = 0) = 1 - p_{\rm a}$, 
where $p_{\rm a} \in (0,1)$ denotes the access probability.
Consequently, with the power control
in \eqref{EQ:Pk}, $x_{k,t} = \sqrt{P_{\rm rx}} d_{k,t} a_k$,
where $a_k \sim {\rm B}(1, p_{\rm a})$. Here,
${\rm B}(n,p)$ 
represents the binomial distribution with parameters $n$ and $p$.
Note that since a high channel gain is a necessary condition for a node
to be active, it is expected that
$$
p_{\rm a} \le \Pr(|h_k|^2 \ge \omega).
$$

At the AP,
CS-based MUD can not only recover the signals from active nodes,
but also identify the active nodes (by estimating $\cI$).
The resulting random access
is referred to as
CRA \cite{Wunder14} \cite{Choi_CRA18} \cite{Choi18b}.
In data-aided sensing,
after the initial round, certain nodes will be asked to 
transmit their measurements by the AP. In this case,
coordinated transmission or controlled access will be used.
However, CRA is still required
due to errors that occur when a request signal is 
sent to nodes, which will be explained in Section~\ref{S:EA}.

\section{A Cross-Layer Design for Data-aided Sensing}	\label{S:DAS}

In the initial round,
the number of active devices, $N$, may not be sufficiently 
large\footnote{Since the sparsity of $\bs$, $S$,
is unknown, the number of active nodes, $N$,
cannot be determined in advance
to guarantee a successful recovery with one round.}
for the AP to have a good estimate of $\bv$.
In this case, another round of sensing/transmissions via CRA
is required, which may result in a different random sensing
of $\bv$. 
In this section, repeated random sensing
is considered
with multiple rounds till a successful recovery of $\bv$,
and data-aided sensing is proposed, which would require
a smaller number of measurements than simple repeated random sensing
by actively choosing nodes for a better reconstruction.

\subsection{Repeated Random Sensing}

Suppose that each node decides to be active or not based on its channel gain 
in each round (due to fading, the channel coefficient
is assumed to be independent in each round).
In this case, random active nodes in each round 
are to transmit their measurements
via CRA, and the AP can acquire the measurements
from them and form $\cD_q$ 
in each round $q$ to estimate $\bv$ 
together with their associated measurement vectors, where
\be
\cD_q = \{m_k, \ k \in \cK_q\},
\ee
which is a subset of $\cD$. Here,
$\cK_q = \cup_{i= 0}^q \cI_i$.
As $q$ increases (under the assumption 
that the nodes transmitted before do not transmit again),
more measurements are available at the AP
(i.e., $\cD_q$ increases), which can lead to a better recovery performance.
To see this clearly,
let $\bw_q$ and $\bB_q$ be the subvector of $\bv$ and 
the submatrix of $\bB$ corresponding to 
$\cI_q$, respectively.
Each round can be seen
as another random projection of $\bs$, and in
round $q$, the AP can have
\be
\left[
\begin{array}{c}
\bw_0 \cr
\vdots \cr
\bw_q \cr
\end{array}
\right] 
= [\bB_0 \ \ldots \ \bB_q]^\rT \bs.
	\label{EQ:wBqs}
\ee
Let $N_q = |\cI_q|$.
While the sparsity of $\bs$ is fixed,
the accumulated number of measurements, i.e., $N_0 + \ldots + N_q$,
increases with $q$. 
Thus, it is expected that the AP is able to reconstruct $\bv$ 
by finding $\bs$ in \eqref{EQ:wBqs}
with a sufficiently large number of rounds
in repeated random sensing.

However,
since the time it takes to complete the reconstruction
as well as the total energy consumption at devices
(to sense and transmit their measurements)
increase with the number of rounds,
it is desirable to have a small
number of rounds if possible.

\subsection{Data-aided Sensing}

In this subsection, 
a method is studied for data-aided sensing
that chooses a set of new measurements
for a better reconstruction based on the existing measurements obtained
from the previous rounds of sensing.

In data-aided sensing, the index set of 
the active nodes in round $q$ is decided
by the data set up to round $q-1$. That is,
$\cI_q$ is given by
\be
\cI_q = \cG_q( \cD_{q-1}),
\ee
where $\cG_q(\cdot)$ is a certain function generating
an index set of the active nodes that potentially
have better measurements for the reconstruction (among the inactive
nodes up to round $q-1$)
based on the data set $\cD_{q-1}$.
Note that in repeated random sensing, 
$\cI_q$ is decided by nodes, which means that
$\cG_q (\cdot)$ is independent of $\cD_{q-1}$.

For convenience, 
$\tilde \cK_q = \cK \setminus \cK_q$,
which is the complement set of $\cK_q$.
Let $\hat \bs_q$ denote the estimate of
$\bs$ after round $q$.
In addition, let
\be
\hat \bv_q = \bB^\rT \hat \bs_q
\ee
be the estimate of $\bv$ after round $q$.
The AP may want a careful selection of active nodes 
for the next round 
to improve recovery performance rather than a random selection.
Let $\bv_q (\cK_{q-1} \cup \cI_q)$ denote the
estimate of $\bv_q$ with data set
associated with
$\cK_{q-1} \cup \cI_q$. Then, for example, the best index set
in round $q$ to minimize 
the squared norm of the error vector 
can be found as
\be
\hat \cI_q = \argmin_{\cI_q \subseteq \tilde \cK_{q-1},
|\cI_q| = N} ||\bv - \bv_q (\cK_{q-1} \cup \cI_q)||^2,
\ee
which is unfortunately impossible to implement
as $\bv$ is not available at the AP.
To avoid this, the Mean Squared Error (MSE) can be employed
as a performance criterion
provided that the statistical properties of $\bv$ is available.
In this paper, however, a different approach that
can be employed without knowing statistical properties of $\bv$
is proposed. 

Consider the following set:
$$
\cI_q = \{k_{q,1}, \ldots, k_{q,N_q}\},
$$
where $k_{q,i}$ is the index of the node that has the $i$th largest
measurement among $\cK \setminus \cK_{q-1}$,
i.e.,
\be
k_{q,i} = \argmax_{k \in \tilde \cK_{q,i}} |v_{k}|^2, \ q \ge 1,
	\label{EQ:ks1}
\ee
where $\tilde \cK_{q,i}$ is the complement set of 
\be
\cK_{q,i}  = \cK_{q-1} \cup \{k_{q,1}, \ldots, k_{q,i-1}\}.
\ee
In \eqref{EQ:ks1}, $\bv$ is required to choose 
the next index set. Since $\bv$ is not available,
$\bv$ can be replaced with its estimate in the previous round,
i.e., $\hat \bv_{q-1}$.
In this case, it can be shown that
\begin{align}
k_{q,i} 
& = \argmax_{k \in \tilde \cK_{q,i}} |[\hat \bv_{q-1}]_{k}|^2 \cr
& = \argmax_{k \in \tilde \cK_{q,i}} |\bb_k^\rT 
\hat \bs_{q-1} |^2, \ q \ge 1,
	\label{EQ:ks2}
\end{align}
which becomes a node selection criterion
for data-aided sensing that depends on
the previous estimate $\hat \bv_{q-1}$.

Suppose that the correlation of
$\bb_k$ and $\bb_{k^\prime}$ is high, where $k \ne k^\prime$.
In this case, it is expected that
$v_k$ and $v_{k^\prime}$ can be highly correlated.
Thus, although the measurement
of node $k$ has a large magnitude,
it may not help improve the performance if $v_{k^\prime}$
is already available at the AP.
This can be taken into account and the 
resulting node section criterion for data-aided sensing
becomes
\be
k_{q,i} = \argmax_{k \in \tilde \cK_{q,i}}
\min_{i \in \cK_{q,i}} 
\frac{ |\bb_k^\rT \hat \bs_{q-1}|^2}{|\bb_k^\rT \bb_i|^2}, \ q \ge 1.
	\label{EQ:ks3}
\ee
In \eqref{EQ:ks3}, $\cI_q$ for round $q$ can be clearly determined
from $\hat \bs_{q-1}$, which is obtained from $\cD_{q-1}$,
for data-aided sensing.

In data-aided sensing,
since the AP chooses active nodes using the node selection
criterion in \eqref{EQ:ks3},
there should be
interactions between connectivity and data layers in an IoT system,
and
controlled access is required to 
acquire measurements from the selected nodes.
As a result, as opposed to repeated random sensing (where there
is no interaction between connectivity and data layers),
data-aided sensing becomes a cross-layer approach.

\section{Controlled Access for Data-aided Sensing}	\label{S:CA}

In this section, an efficient 
controlled access scheme with low signaling overhead
is discussed for data-aided sensing 
by exploiting
unique signature sequences of nodes.

\subsection{Compressive Transmission Request}

For controlled access, certain request signals
are to be sent to nodes from the AP
in data-aided sensing, which are not necessary in
repeated random sensing. Thus, if the length of request signals
is long, it may offset the advantage of 
data-aided sensing over 
repeated random sensing. To avoid it,
a simple but efficient approach 
is developed using the structure of CRA
considered in Subsection~\ref{SS:CRA}.

Prior to round $q$,
the AP can request the nodes selected by \eqref{EQ:ks2}
to transmit their measurements, i.e., it 
can send a request message to each node
in $\cI_q$ at a time. However, the total time 
to request becomes proportional to $|\cI_q|$.
To shorten the time to request, the broadcast nature 
of the wireless medium and the structure of CRA
can be exploited
so that one request signal might be sufficient.
To this end, the following downlink probe or
request signal can be transmitted from the AP to the nodes:
\be
\bp_q = \sum_{k \in \cI_q} \bc_k.
\ee
For simplicity, it is assumed that $N_q = |\cI_q| 
= N$ for all $q$ in the rest of this
paper.
The channel coefficient from the AP to node
$k$ becomes $h_k$ based on  the channel reciprocity in TDD
and the received (compressive
transmission) request signal becomes
\be
\by_k = h_k \frac{A_{\rm ap}}{||\bp_q||} \bp_q  + \bn_k,
\ee
where $n_k \sim \cC \cN (0, N_0 \bI)$ is the background noise vector
at node $k$ and $A_{\rm ap}$ is the 
amplitude of the request signal from the AP.
The resulting request signal is referred to as compressive
transmission request signal.

\subsection{Hypothesis Testing for Request Signal Detection}

Since nodes are requested by
the AP to transmit their measurements from round 1 in data-aided sensing,
each node (not yet transmitted) needs to perform hypothesis testing 
with the received request signal, $\by_k$, 
to see whether or not it is requested.
In this subsection,
hypothesis testing is studied
using the Log-Likelihood Ratio (LLR) 
\cite{ChoiJBook2} and find the error probabilities.

At node $k$, in order
to see whether or not it is requested to transmit its measurement,
it can use the following output of the correlator with its
signature sequence as a test statistic:
\begin{align}
z_k = \Re \left((h_k \bc_k)^\rH \by_k \right) 
= 
|h_k|^2 \frac{A_{\rm ap}}{||\bp_q||}
\indicator (k \in \cI_q) + \Re\left( h_k^* w_k \right),
\end{align}
where $\indicator(\cdot)$ represents the indicator function
and
$w_k$ is the interference-plus-noise vector.
Suppose that each element of $\bc_k$ is random 
and is one of $\{(\pm 1 \pm j)/\sqrt{2L}\}$.
Then, $\bc_k^\rH \bc_i$, $i \ne k$, 
can be approximated as a
Gaussian random variable (thanks to  the central limit
theorem) with zero mean and variance $\frac{1}{L}$.
In addition, it is assumed that
$||\bp_q||^2 \approx N$
by approximating $\bc_k^\rH \bc_i \approx 0$
for $k \ne i$
(which might be reasonable when $L$ is sufficiently large).
Let $\cH_1$ and $\cH_0$ denote the hypotheses
of $k \in \cI_q$ and $k \notin \cI_q$, respectively.
Then, the Probability Density
Function (pdf) of $z_k$ under hypothesis $p$,
denoted by $f_p (z_k)$, $p \in \{0,1\}$,
is approximately given by
\begin{align}
f_0(z_k) & = \cN\left(0, \frac{|h_k|^2 \sigma_{w,0}^2}{2} \right) \cr
f_1(z_k) & = \cN\left(\sqrt{P_{\rm ap}}|h_k|^2, 
\frac{|h_k|^2 \sigma_{w,1}^2}{2} \right),
	\label{EQ:ff01}
\end{align}
where $P_{\rm ap} = \frac{A^2_{\rm ap}}{N}$ and
\begin{align}
\sigma_{w,1}^2 & = \uE[|w_k|^2] 
= \uE 
\left[|\bc_k^\rH 
\left(\sum_{i \in \cI_q \setminus k} \bc_i + \bn_k \right) |^2 \right] \cr
& = P_{\rm ap} \frac{N - 1}{L} + N_0 \cr
\sigma_{w,0}^2 & = P_{\rm ap} \frac{N}{L} + N_0.
\end{align}
A device that did not transmit yet will have a higher probability
to be requested as the number of rounds increases,
and the a priori probabilities in round $q$ becomes
$P_q(\cH_0) = \frac{K-qN}{K - (q-1)N}$ and
$P_q(\cH_1) = \frac{N}{K - (q-1)N}$.
Here, $K - (q-1)N$ is the number of the devices 
that did not transmit their measurements until round $q$.

If node $k$ does not transmit its measurement up to round $q-1$,
in round $q$,
it can consider the following LLR test:
\be
U_k = \ln \frac{f_1 (z_k)}{f_0 (z_k)} \defh 
\tau_q,
	\label{EQ:LT}
\ee
where $\tau_q$ is a threshold.
Letting $\tau_q =
\ln \frac{P_q (\cH_0)}{P_q (\cH_1)}
= \ln \frac{K  - qN}{N}$,
the resulting test is based on the Maximum A posteriori Probability
(MAP) criterion.
When $L$ is sufficiently large, i.e.,
$\frac{1}{L} \ll 1$, it can be shown that
$\sigma_{w,0}^2 \approx \sigma_{w,1}^2$
and the test in \eqref{EQ:LT} can be approximated as follows:
\be
z_k \defh \frac{\sqrt{P_{\rm ap}} |h_k|^2}{2} 
+ \frac{\sigma_{w,0}^2 \tau_q}{2 \sqrt{P_{\rm ap}} }.
\ee
Let $\gamma = \frac{P_{\rm ap}}{\sigma_{w,0}^2}$,
which is the downlink signal-to-interference-plus-noise ratio (SINR).
If $\tau_q = |h_k|^2 \gamma u_q$,
it can be shown that
\be
z_k \defh 
\frac{\sqrt{P_{\rm ap}} |h_k|^2}{2} (1  + u_q 
\gamma).
	\label{EQ:z_g}
\ee
Here, the design parameter, $\gamma u_q$, which is referred to as
the scaled decision parameter for convenience,
will be discussed in Section~\ref{S:EA}.
Note that since the truncated channel inversion power control policy 
is considered in this paper,
as in \eqref{EQ:homega}, if the channel gain is weak,
the node cannot respond the request although it can 
correctly detect the request signal.

A reliable approach can be adopted to deliver the request
signal to nodes.
For example, a Hybrid Automatic
Repeat reQuest (HARQ) protocol \cite{WickerBook}
can be used to deliver a request signal through a dedicated
downlink channel. In this case, the signaling overhead becomes high
although the request signal can be reliably delivered (with 
a very low error probability).
On the other hand, the compressive transmission request 
approach (which is aimed at low signaling overhead)
in this section is not sufficiently reliable 
and results in an erroneous decision at a node 
with a relatively high probability.
However, even if a node that is not requested sends 
its measurement, the AP can still utilize this information. 
Thus, erroneous decisions on the request signal at nodes 
may have little effect on the overall performance of data-aided sensing,
which justifies using
compressive transmission request with
low signaling overhead.

\section{Error Analysis}	\label{S:EA}

In this section, possible 
erroneous decisions at a node 
are studied
when the compressive transmission request
is used and their impact on the performance. 

\subsection{Decision Errors in Downlink}

In round $q \ (\ge 1)$
(i.e., after the initial round),
a certain node can be requested to transmit its measurement
in data-aided sensing. However,
the node may not correctly detect this request
signal
or the required transmit power can be higher than $P_{\rm max}$
due to fading, which makes
the node unable to respond in round $q$.
This results in a Missed Detection (MD) event in downlink.
The probability of MD for $k \in \cI_q$ becomes
\begin{align}
P_{{\rm MD},q} 
& = \Pr( U_k < \tau_q  , |h_k|^2 \ge \omega\,|\, k \in \cI_q) \cr
& \quad + \Pr(|h_k|^2 < \omega \,|\, k \in \cI_q).
	\label{EQ:PMD}
\end{align}
Furthermore, node $k \notin \cI_q$ happens to transmit its
measurement although it is not asked with the following
probability:
\begin{align}
P_{{\rm FA},q} = \Pr( U_k \ge \tau_q  , |h_k|^2 \ge \omega \,|\, k 
\notin \cI_q),
\end{align}
which is 
the probability of False Alarm (FA).
Due to the events of MD and FA,
which are referred to as downlink (decision) errors for
convenience,
the index set of active nodes in round $q$
can be different from $\cI_q$.
Thus, it is necessary for the AP to perform CS-based MUD
as in round 0 to not only decode the signals, but also
identify active nodes.

When $L$ is sufficiently large,
\eqref{EQ:z_g} can be used to find the probability of FA.
From the pdf of $z_k$ under $\cH_0$ in \eqref{EQ:ff01},
it can be shown that
\begin{align}
& \Pr \left(z_k \ge 
\frac{\sqrt{P_{\rm ap}} |h_k|^2}{2} (1  + u_q 
\gamma) \,|\, \cH_0, h_k \right) \cr
& = 
\int_{\frac{\sqrt{P_{\rm ap}} |h_k| (1+ u_q \gamma)}{2}}^\infty
\frac{1}{\sqrt{\pi |h_k|^2 \sigma_{w,0}^2}}
e^{-\frac{z_k^2}{|h_k|^2 \sigma_{w,0}^2}}
d z_k \cr
& = 
\cQ \left(
\frac{|h_k|}{\sqrt{2}} \sqrt{\gamma }( 1+ u_q \gamma)
\right).
\end{align}
Note that for a Rayleigh channel,
$|h_k|^2$ has the following exponential pdf:
\be
G = |h_k|^2 \sim f_G (g) = \exp(- g), \ g \ge 0,
	\label{EQ:pdfW}
\ee
under the assumption that the channel power gain 
is normalized, i.e., $\uE[|h_k|^2] = 1$.
Then, the average probability of FA can be approximately obtained
as
\begin{align}
P_{{\rm FA}, q} 
& \approx \int_\omega^\infty
\cQ \left(
\sqrt\frac{g}{2} 
\sqrt{\gamma }( 1+ u_q \gamma)
\right) f_G (g) d g \cr
& \approx 
\int_\omega^\infty 
\left(
\frac{ e^{- \frac{g\gamma (1+ u_q \gamma)^2} {4} }}{12}
+
\frac{ e^{- \frac{g\gamma (1+ u_q \gamma)^2} {3} }}{4}
\right)
e^{-g} d g \cr
& =  \frac{e^{ - \omega \left(1+ \frac{\gamma}{4}(1 + u_q \gamma)^2 \right) }
}{12 (1 + \frac{\gamma}{4} (1+ u_q \gamma)^2)}
+ \frac{e^{ - \omega \left(1+ \frac{\gamma}{3}(1 + u_q \gamma)^2 \right) }
}{4 (1 + \frac{\gamma}{3} (1+ u_q \gamma)^2)} , \ \ 
	\label{EQ:APFA}
\end{align}
where the second approximation is due to the following approximation of
the Q-function in \cite{Chiani02}:
$$
\cQ(x) \approx 
\frac{1}{12} e^{-\frac{x^2}{2}}
+
\frac{1}{4} e^{-\frac{2x^2}{3}}.
$$

Similarly, after some manipulations, it can be shown that
\begin{align}
P_{{\rm MD},q} 
& \approx \int_\omega^\infty
\cQ \left(
\sqrt{\frac{g}{2}}
\sqrt{\gamma } ( 1 - u_q \gamma) \right) f_G (g) dg \cr
& + \int_0^\omega f_G (g) dg \cr
& \approx
\frac{e^{ - \omega \left(1+ \frac{\gamma}{4}(1 - u_q \gamma)^2 \right) }
}{12 (1 + \frac{\gamma}{4} (1- u_q \gamma)^2)} +
\frac{e^{ - \omega \left(1+ \frac{\gamma}{3}(1 - u_q \gamma)^2 \right) }
}{4 (1 + \frac{\gamma}{3} (1- u_q \gamma)^2)} \cr
& \quad + 1 - e^{-\omega}.
	\label{EQ:APMD}
\end{align}
In addition, as shown in \eqref{EQ:PMD},
the probability of MD is lower-bounded by
$\Pr(|h_k|^2 < \omega \,|\, k \in \cI_q)
= \Pr(|h_k|^2 < \omega) = 1 - e^{-\omega}$,
which is independent of $\gamma$.
Thus, although $\gamma \to \infty$,
the average probability of MD cannot be low,
while the average probability of FA approaches 0.

Note that from \eqref{EQ:APFA} and 
\eqref{EQ:APMD}, it can be shown that the probability of MD increases and
that of FA decreases with $\gamma u_q$.
In addition, to keep both the error probabilities reasonably low,
it is required that
$1 + u_q \gamma > 0$ and $1 - u_q \gamma > 0$,
or
\be
-1 < u_q \gamma < 1,
\ee
i.e., the scaled decision parameter, $u_q \gamma$, has to be bounded
between $-1$ and $1$.


\subsection{CRA and Design Issues}

According to \eqref{EQ:APMD} and \eqref{EQ:APFA}, 
there might be downlink errors (i.e., MD and FA events)
and the AP expects to see that the actual index set of active
nodes, which is denoted by $\tilde \cI_q$, 
in round $q \ge 1$ may differ from $\cI_q$.
From this, even in data-aided sensing,
CRA plays a key role in receiving measurements from 
active nodes, because
the AP still needs to perform the activity detection
to estimate $\tilde \cI_q$.

The performance of CRA
is studied in \cite{Choi_CRA18} based on the notion of
multiple measurement vectors (MMV) \cite{Chen06} \cite{Davies12}.
As in \eqref{EQ:br},
for a sufficiently long length of packet,
the notion of MMV can be applied to the activity detection
at the AP. Under certain conditions, it is possible 
to detect up to $L-1$ signals \cite{Chen06} \cite{Davies12}.
In practice, however, the number of the signals that can be
detected is smaller than $L-1$ due to the background noise
as well as the limited complexity of the receiver algorithm used.
For convenience, let $\bar N$ be the maximum number of the signals
that can be detected at the AP
with a high probability, where $\bar N \le L-1$.
In this case, 
$N_q \le \bar N$ is required
so that the AP is able to decode the signals
from active nodes with a high probability.

However, although $|\cI_q| = N_q \le \bar N$ at the AP,
the actual number of the active nodes in round $q \ge 1$, which 
is $\tilde N_q = |\tilde \cI_q|$, can be different from $N_q$. 
Thus, the scaled decision parameter $\gamma u_q$ at each node
needs to be carefully decided to make sure that $\uE[\tilde N_q]
\le \bar N$. In the presence of downlink errors, since
$N_q$ is no longer the number of active nodes,
it is referred to
as the required number of active nodes in round $q$.

While $\gamma u_q$ can be decided to meet $\uE[\tilde N_q] \le \bar N$,
there is also another important issue to be taken into account.
The performance of data-aided sensing 
depends on the probability of MD. 
If the AP cannot have the measurements from the selected nodes
in round $q \ge 1$ due to MD events,
it cannot provide a good estimate of $\bv$.
Thus, it is desirable to have a low probability of MD.
As shown in \eqref{EQ:APMD},
in order to have a low probability of MD,
it is  required that the scaled decision parameter is less than 1
(i.e., $u_q \gamma < 1$).
However, unfortunately, 
when the probability of MD decreases,
the probability of FA increases
and  $\tilde N_q$ increases.
To see this clearly, consider
the average number of active nodes in round $q$ as follows:
\begin{align}
\uE[\tilde  N_q] 
& = N_q (1 - P_{{\rm MD}, q}) \cr
& + (K - 
(\tilde N_0 +
\tilde N_1+ \ldots + \tilde N_{q-1} + N_q)) P_{{\rm FA}, q} \cr
& \le N_q (1 - P_{{\rm MD}, q}) 
+ K P_{{\rm FA}, q}.
	\label{EQ:EtN}
\end{align}
From this, the increase of $P_{{\rm FA}, q}$ 
(by decreasing $P_{{\rm MD}, q}$)
results in the increase of $\bar \uE[\tilde N_q]$.
Thus, in order to avoid unsuccessful decoding in CRA,
a low probability of FA is needed,
which however leads to a high  probability of MD
and degrades the performance of data-aided sensing.
This shows that the scaled decision parameter,
$u_q \gamma$, becomes a key parameter to be carefully decided.
That is, as $u_q \gamma$ decreases (or $u_q \gamma$ approaches $-1$),
the number of FA events increases, which leads to unsuccessful CRA.
On the other hand, as $u_q \gamma$ increases (or $u_q \gamma$ approaches $1$),
the number of MD events increases, 
offsetting the benefit of data-aided sensing.

\section{Simulation Results}	\label{S:Sim}

In this section,
simulation results for data-aided sensing
are presented. For simulations, it is 
assumed that $\bB^\rT$ consists of randomly selected
$M$ column vectors of the $K \times K$ Discrete Cosine Transform (DCT) 
and the non-zero $S$ elements of the sparse vector $\bs$ are either 
$+1$ or $-1$ (with the equal probability).
To reconstruct $\bv$ at the AP
with a subset of the measurements sent by active nodes,
the lasso method \cite{Tibshirani96} \cite{Hastie15}
is used. That is,
in each round $q$, 
with the data set of measurements, $\cD_q$,
the sparse signal, $\hat \bs_q$, is estimated using the lasso method.
Note that for the sake of simplicity\footnote{The performance of CRA
also depends on the algorithm used for CS-based MUD. 
To avoid this dependency, an ideal performance
of CRA is simply considered, which can detect up to $L-1$ signals. 
An algorithm of reasonably complexity can have a near
ideal performance as studied in \cite{Choi_CRA18}.}, 
CRA is not considered
in simulations. However, as shown in \cite{Choi_CRA18},
as long as the actual number of active nodes
is sufficiently smaller than the length of signature sequences, $L$, 
CRA becomes successful. That is,
in controlled access, it is assumed that the AP is able to decode
signals from the requested nodes as well as those from 
other nodes that happen to be active (due to FA events)
provided that $\tilde N_q \le L-1$.
In addition, it is assumed that $N_q = N$ for all $q$.

\begin{figure}[thb]
\begin{center}
\includegraphics[width=\figwidth]{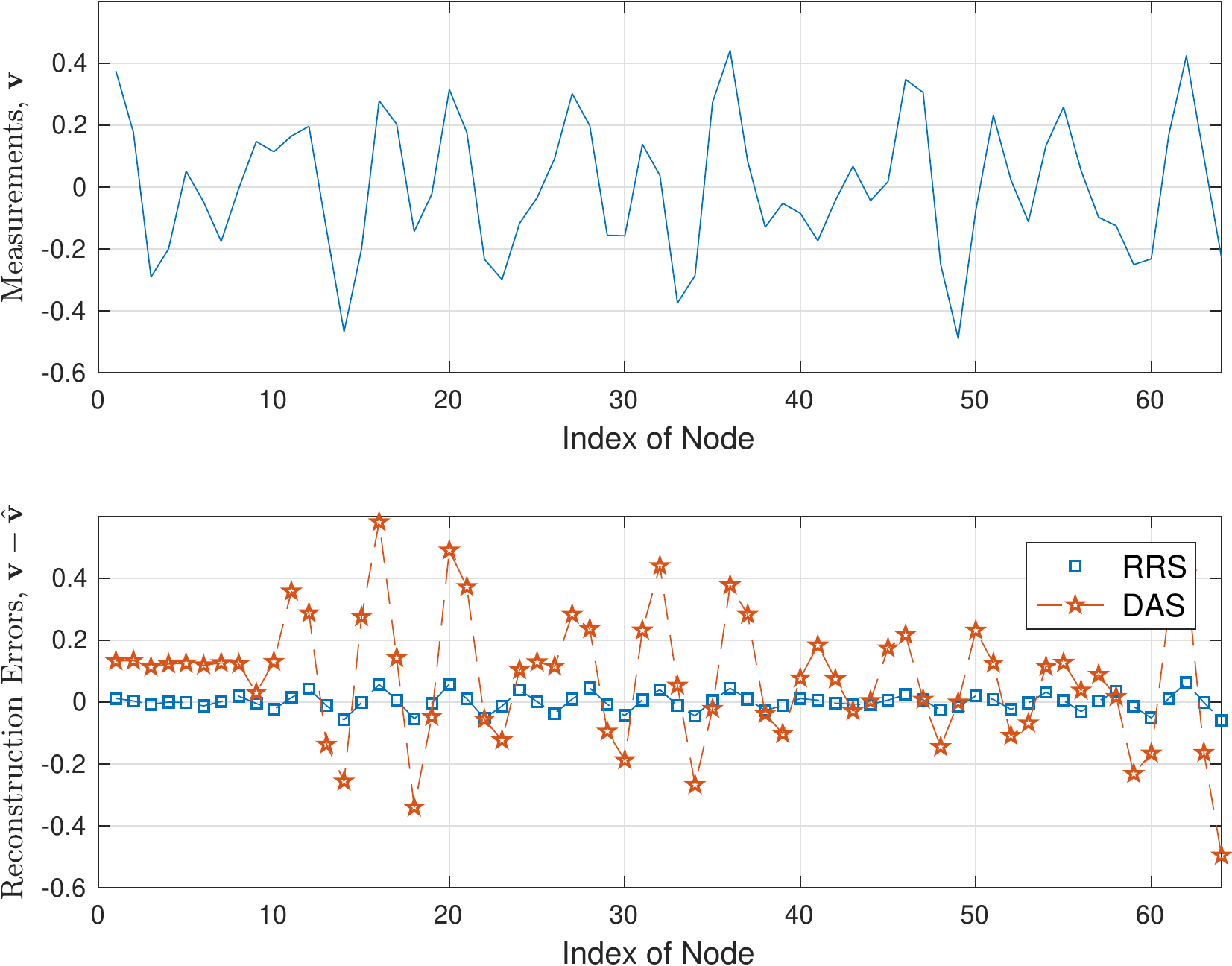}
\end{center}
\caption{The measurements at $K$ nodes (the upper plot)
and the reconstruction errors after $4$ rounds
(the lower plot)
when $K = 64$, $M = 25$, $S = 3$, and $N = 5$.}
        \label{Fig:plt_wav}
\end{figure}

Fig.~\ref{Fig:plt_wav} shows
reconstruction errors at the AP with data-aided sensing
and repeated random sensing after 4 rounds together with
the target signal, $\bv$,
when $K = 64$, $M = 25$, $S = 3$, and $N = 5$.
In the legend, RRS and DAS stand for
the results of repeated random sensing and 
data-aided sensing, respectively.
As shown in Fig.~\ref{Fig:plt_wav},
data-aided sensing can provide a good estimate
of $\bv$ at the AP with $4 N = 20$ measurements,
while repeated random sensing cannot provide a reasonable estimate.

As mentioned earlier, 
the AP needs to choose active nodes using
controlled access in data-aided sensing. However,
due to
downlink errors (i.e., MD and FA events), the AP may not be able to have
the measurements from the selected nodes in $\cI_q$, $q \ge 1$,
which can degrade the performance.
In particular, if there are a number of FA events,
the AP can have more measurements from 
the nodes that are not requested to transmit than those from the nodes
that are requested.
Thus, it is important to take into account MD and FA events
in compressive transmit request.
In the rest of simulation results, the values of 
$\omega = \frac{P_{\rm rx}}{P_{\rm max}}$ and $\frac{A_{\rm ap}^2}{N_0}$
are set
to ${\text -}10$ dB and $20$ dB, respectively.
In addition, $M = 100$ and $L = 64$ are fixed
and a Rayleigh fading model in \eqref{EQ:pdfW}
is considered for $h_k$ in compressive transmission request.

\begin{figure}[thb]
\begin{center}
\includegraphics[width=\figwidth]{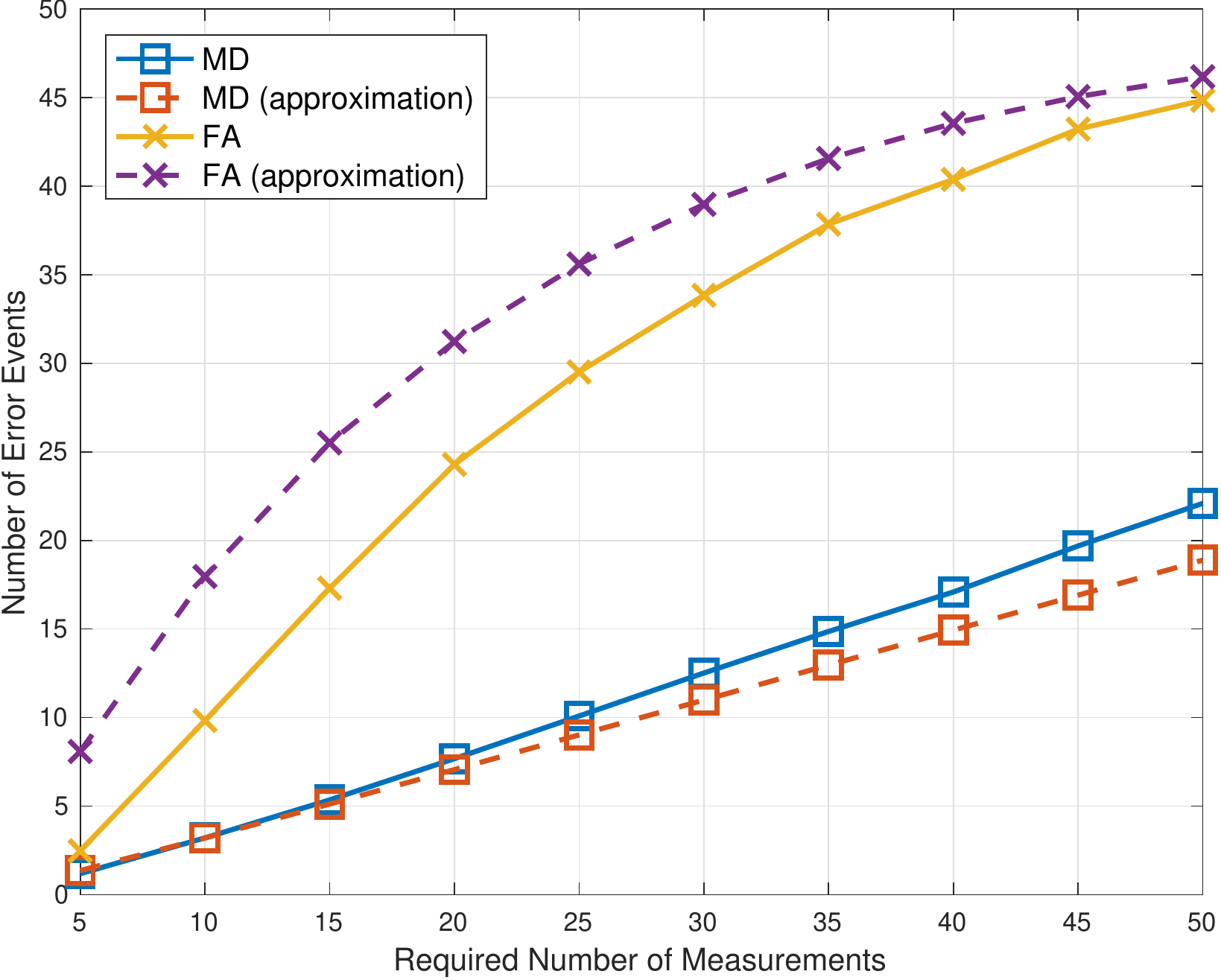}
\end{center}
\caption{The (average) numbers of MD and FA events (from 1000 runs)
as functions of the requested number of measurements, $N$,
when $K = 300$, $L = 64$, and $\gamma u_q = 0.5$.}
        \label{Fig:plt_d1}
\end{figure}

Fig.~\ref{Fig:plt_d1}
shows the (average) numbers of MD and FA events
(from 1000 runs)
as functions of the requested number of measurements, $N$,
when $K = 300$, $L = 64$, and $\gamma u_q = 0.5$. 
For the theoretical approximations of the average number of MD and FA, 
\eqref{EQ:APMD} and \eqref{EQ:APFA} 
are used, respectively,
in Fig.~\ref{Fig:plt_d1}.
It is shown that as $N$ increases, there are more FA events
than MD events. For example, if $N = 50$,
there are about 45 FA events and 20 MD events. This implies
that the AP can only have 30 desired measurements,
while 45 measurements are transmitted from the nodes
that are not requested due to FA events.
More importantly, the actual number of active nodes
becomes $30 + 45 = 75$, which is greater than $L -1 = 63$.
Therefore, when $N$ is too large, the AP not only has more undesirable
measurements, but also fails to perform CRA successfully.
This implies that $N$ should be carefully decided
so that $\uE[\tilde N_q]$ is smaller than $L-1$ 
to avoid at least unsuccessful CRA.

\begin{figure}[thb]
\begin{center}
\includegraphics[width=\figwidth]{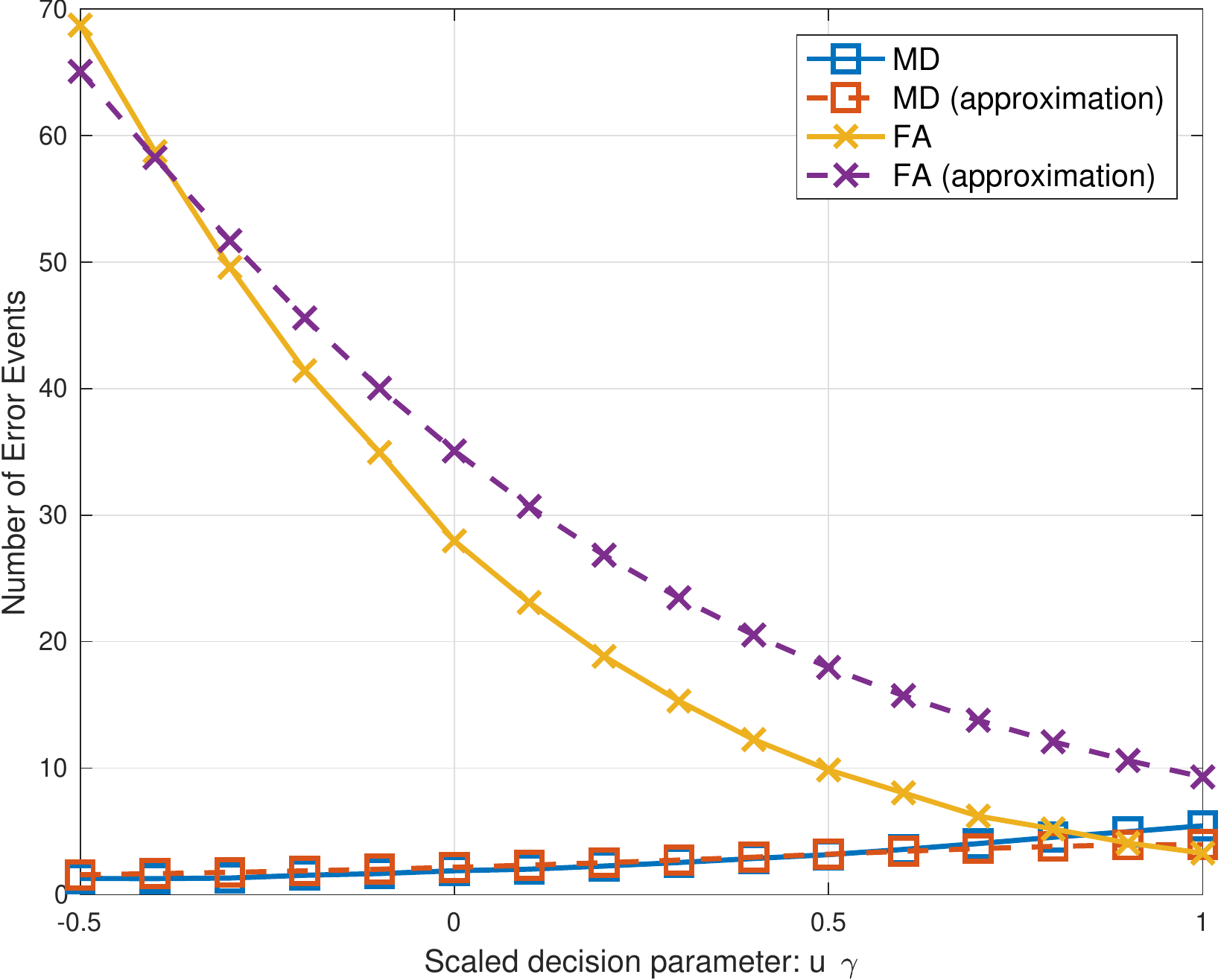} 
\end{center}
\caption{The 
(average) numbers of MD and FA events (from 1000 runs)
as functions of $\gamma u_q = 0.5$
when $K = 300$, $L = 64$, and $N = 10$.}
        \label{Fig:plt_d2}
\end{figure}

Fig.~\ref{Fig:plt_d2} shows the 
(average) numbers of MD and FA events (from 1000 runs)
as functions of the scaled decision parameter, $\gamma u_q$,
when $K = 300$, $L = 64$, and $N = 10$.
As expected, the number of MD increases and
that of FA decreases when $\gamma u_q$ increases.
As mentioned earlier, a large $\gamma u_q$ 
is required to keep the actual number of active nodes
small so that CRA becomes successful.
On the other hand, the benefit of data-aided sensing will diminish
if $\gamma u_q$ is too large (due to a large number of MD events).

\begin{figure}[thb]
\begin{center}
\includegraphics[width=\figwidth]{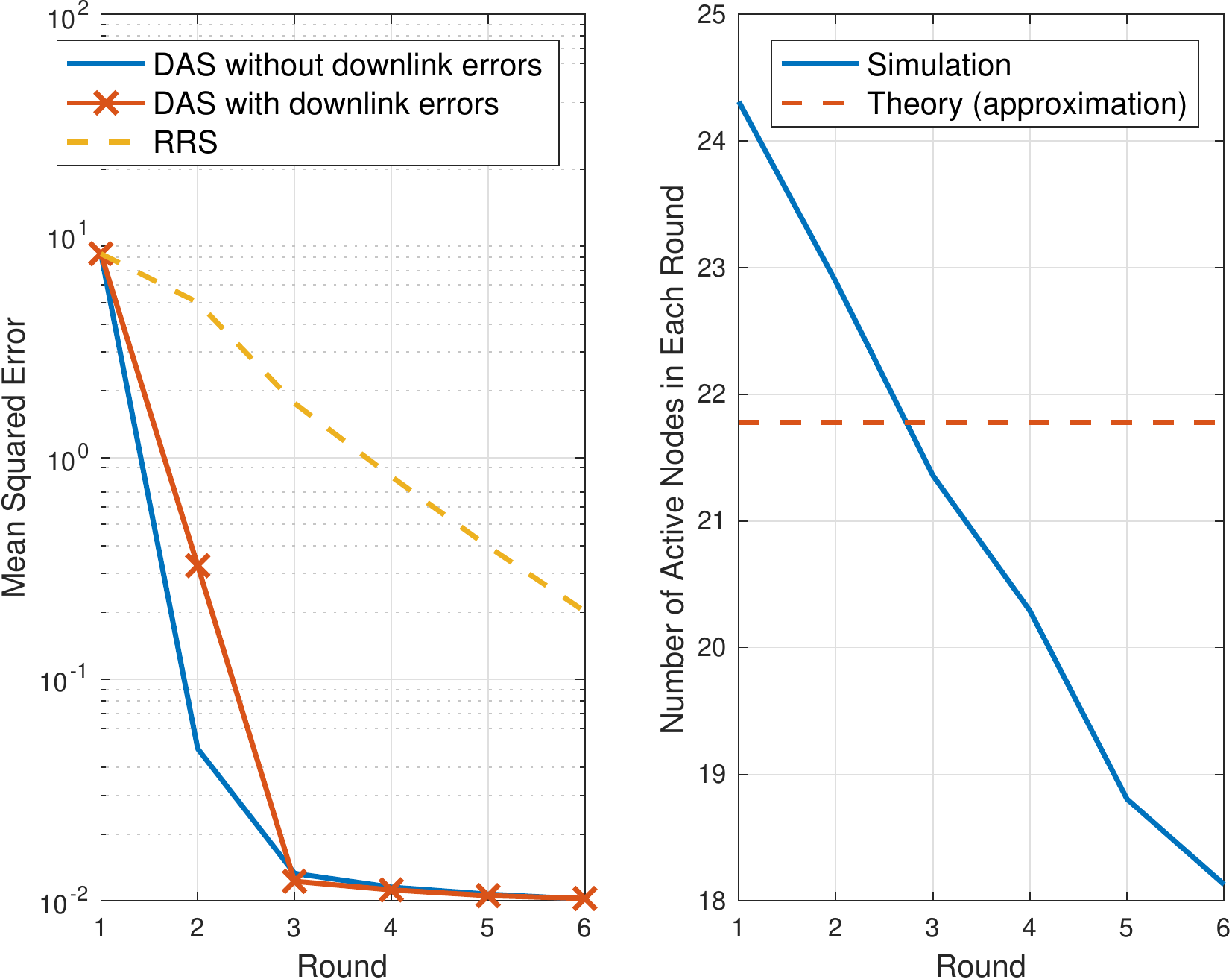} \\
\hskip 0.5cm (a) \hskip 3.5cm (b)
\end{center}
\caption{Performances of data-aided sensing and 
repeated random sensing as functions of rounds, $q$
when $K = 300$, $N = 10$, 
$\gamma u_q = 0.5$ (for all $q$), and $S = 10$:
(a) mean squared error, $\uE[||\bv - \bv_q||^2]$;
(b) the number of active nodes in each round.}
        \label{Fig:plt1}
\end{figure}

With downlink errors 
on compressive transmission request
(i.e., FA and MD events),
the performance of data-aided sensing
is shown in Fig.~\ref{Fig:plt1},
where the mean squared error, $\uE[||\bv - \bv_q||^2]$,
is shown in each round
together with the number of active nodes in each round
when $K = 300$, $N = 10$, and $S = 10$.
For fair comparisons, the actual number of active
nodes in each 
round, $\tilde N_q$, is assumed to be the same in data-aided
sensing (with and without downlink errors) and repeated random sensing.
To predict the number of active nodes,
the upper-bound in \eqref{EQ:EtN} is considered.
In Fig.~\ref{Fig:plt1} (a),
it is shown that downlink errors
degrade the performance of data-aided sensing,
which is however still better than the performance of
repeated random sensing.
In Fig.~\ref{Fig:plt1} (b),
the actual number of active
nodes from simulations
decreases as the number of round, $q$, increases
However, the approximation from the theory 
remains unchanged as the upper-bound (i.e., \eqref{EQ:EtN})
is used with fixed $u_q \gamma = 0.5$ and $N = 10$ for all $q$.
Note that since the actual number of active
nodes is smaller than $L -1 = 63$, successful
CRA in controlled access is 
assumed.

\begin{figure}[thb]
\begin{center}
\includegraphics[width=\figwidth]{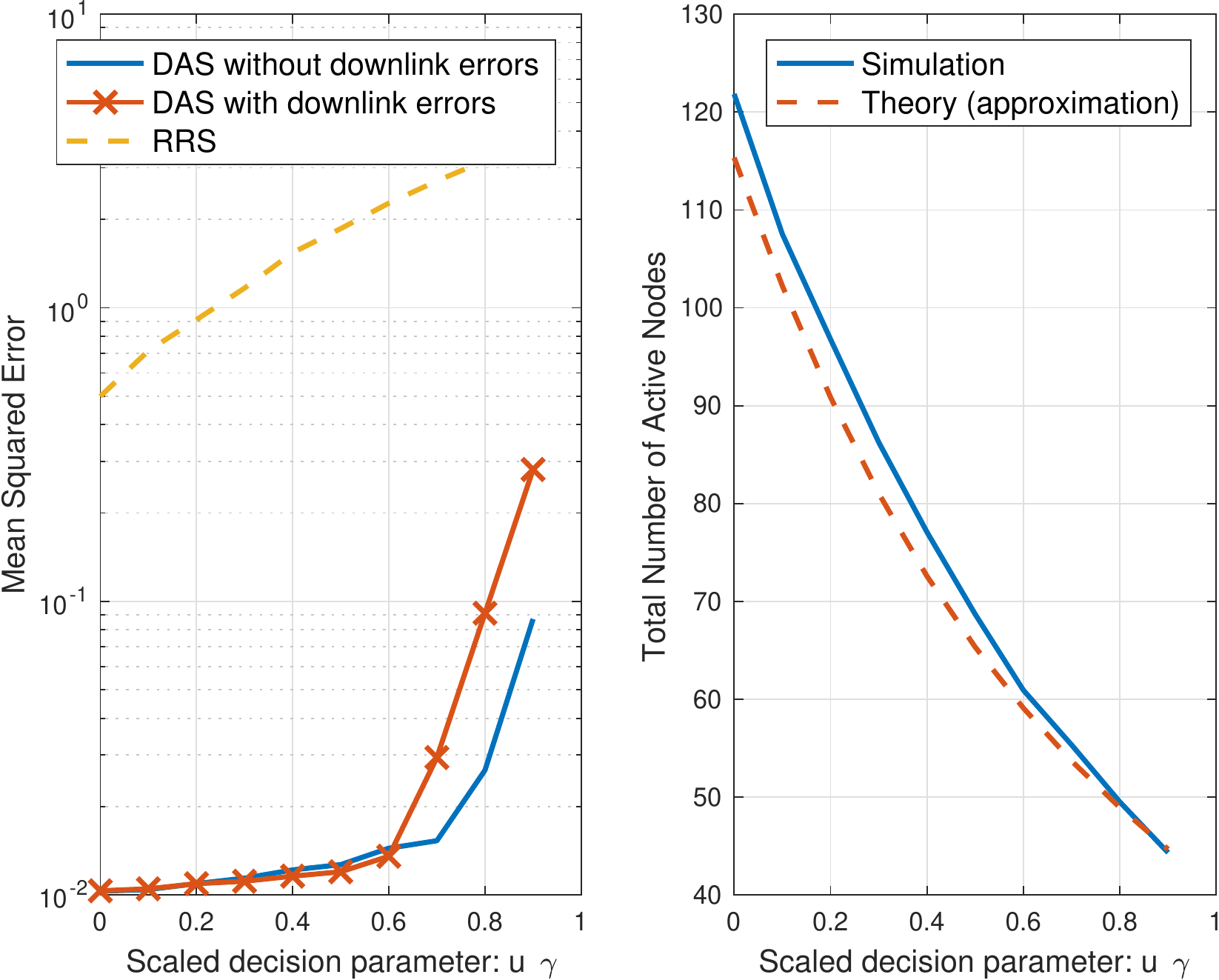} \\
\hskip 0.5cm (a) \hskip 3.5cm (b)
\end{center}
\caption{Performances of data-aided sensing and 
repeated random sensing for different values of $\gamma u_q$ 
when $K = 300$, $N = 10$ (for all $q$), 
and $S = 10$ (with a fixed number of rounds, $q = 3$):
(a) mean squared error, $\uE[||\bv - \bv_q||^2]$ with $q = 3$;
(b) the average total number of active nodes,
$\uE[\tilde N_0 + \ldots + \tilde N_q]$.}
        \label{Fig:plt2}
\end{figure}

Fig.~\ref{Fig:plt2} shows
the performances of data-aided sensing and 
repeated random sensing for different values of $\gamma u_q$ 
when $K = 300$, $N = 10$,
and $S = 10$ (with a fixed number of rounds, $q = 3$).
As expected, as the scaled decision parameter, $\gamma u_q$,
increases, the actual number of active nodes
(or the number of measurements at the AP)
decreases as shown in Fig.~\ref{Fig:plt2} (b),
which results in the increase of MSE in Fig.~\ref{Fig:plt2} (a).
Thus, for a reasonable performance, $\gamma u_q$ is to be carefully
chosen (in this case, $\gamma u_q$ can be less than $0.6$
as long as the actual number of active nodes per round
is less than $L-1$ for successful CRA).

\begin{figure}[thb]
\begin{center}
\includegraphics[width=\figwidth]{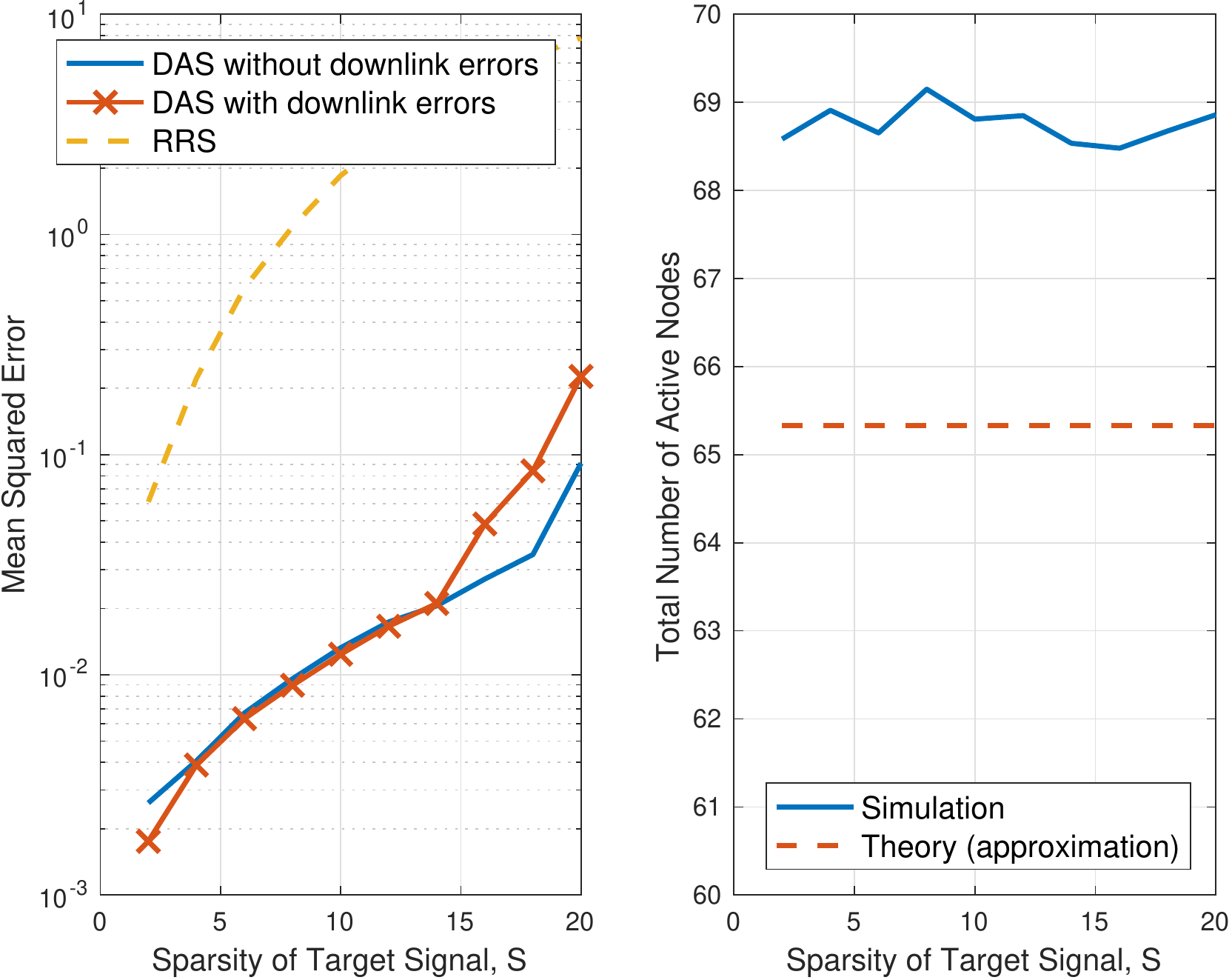} \\
\hskip 0.5cm (a) \hskip 3.5cm (b)
\end{center}
\caption{Performances of data-aided sensing and 
repeated random sensing for different numbers of 
$S$ when $K = 300$, $N = 10$,
and $\gamma u_q = 0.5$ (for all $q$), 
(with a fixed number of rounds, $q = 3$):
(a) mean squared error, $\uE[||\bv - \bv_q||^2]$ with $q = 3$;
(b) the average total number of active nodes,
$\uE[\tilde N_0 + \ldots + \tilde N_q]$.}
        \label{Fig:plt3}
\end{figure}

In Fig.~\ref{Fig:plt3},
the performances of data-aided sensing and 
repeated random sensing are shown for different numbers of 
the sparsity, $S$, when $K = 300$, $N = 10$,
and $\gamma u_q = 0.5$ (for all $q$)
with a fixed number of rounds, $q = 3$.
Clearly, as the sparsity decreases, the AP can have a lower MSE.
It is noteworthy that 
the MSE of data-aided sensing with downlink errors
(i.e., MD and FA events) can be lower than
that without downlink errors as shown in 
Fig.~\ref{Fig:plt3} (a) when $S$ is small.
That is, the measurements from the set of the nodes that are selected
by the node selection criterion in \eqref{EQ:ks3} can have 
a higher MSE than the measurements from a slightly different set
of nodes.
This demonstrates that \eqref{EQ:ks3} 
does not exactly lead to the decrease
of MSE (i.e., \eqref{EQ:ks3} is not optimal in terms of the MSE),
while it is a reasonable criterion to improve
the performance in general.
As mentioned earlier, if 
the AP can have statistical properties of $\bv$, it can afford to find
an MSE-based criterion,
while \eqref{EQ:ks3} 
is applicable to any $\bv$ that has a sparse 
representation (without any knowledge of statistical properties of 
$\bv$).

\begin{figure}[thb]
\begin{center}
\includegraphics[width=\figwidth]{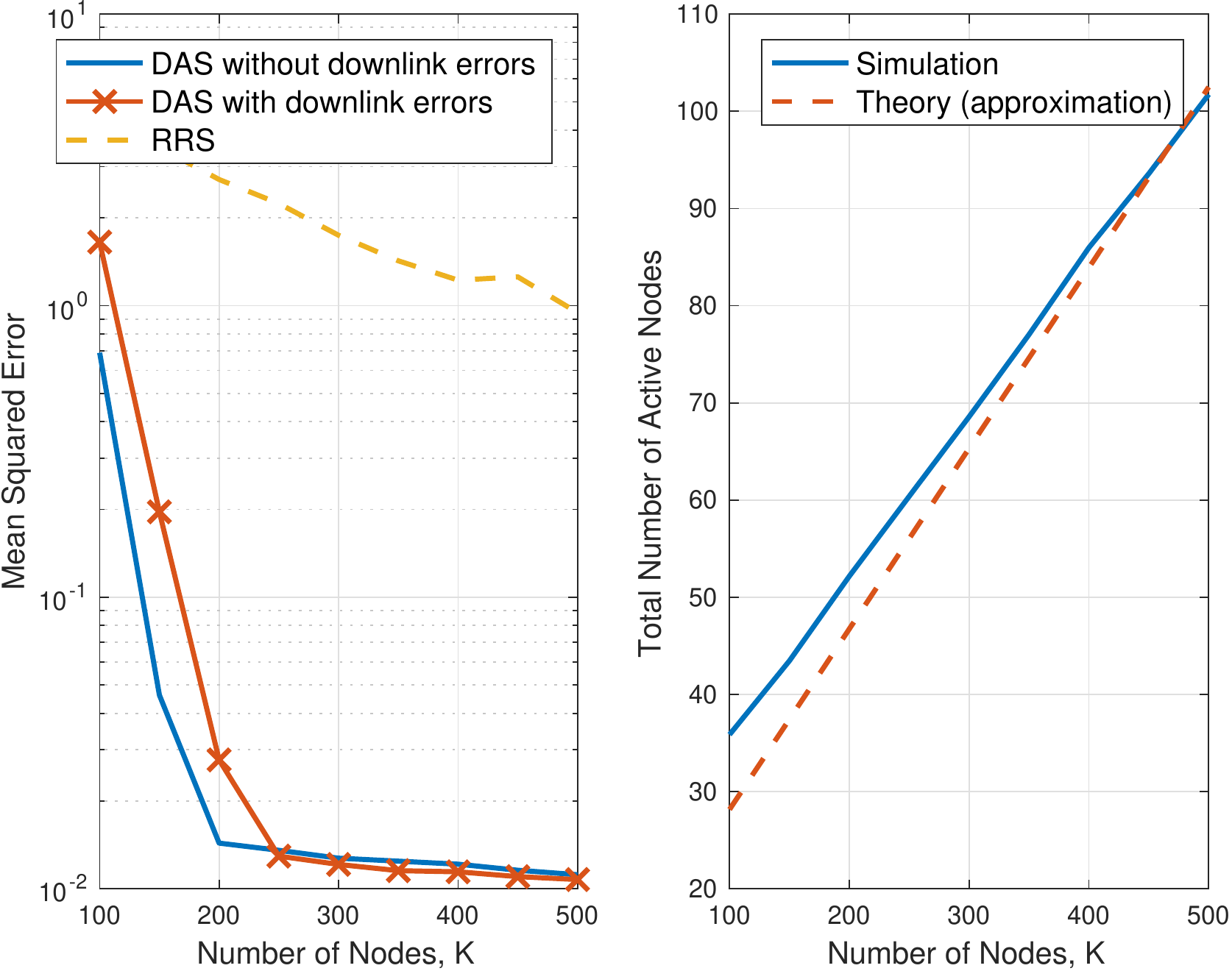} \\
\hskip 0.5cm (a) \hskip 3.5cm (b)
\end{center}
\caption{Performances of data-aided sensing and 
repeated random sensing for different numbers of nodes,
$K$, when $N = 10$, $\gamma u_q = 0.5$ (for all $q$), 
and $S = 10$ (with a fixed number of rounds, $q = 3$):
(a) mean squared error, $\uE[||\bv - \bv_q||^2]$ with $q = 3$;
(b) the average total number of active nodes,
$\uE[\tilde N_0 + \ldots + \tilde N_q]$.}
        \label{Fig:plt4}
\end{figure}

Fig.~\ref{Fig:plt4} shows
the performances of data-aided sensing and 
repeated random sensing for different numbers of 
nodes, $K$, when $N = 10$, $\gamma u_q = 0.5$ (for all $q$), 
and $S = 10$ (with a fixed number of rounds, $q = 3$).
Since the actual number of active nodes
increases with $K$ as shown in Fig.~\ref{Fig:plt4} (b)
although $N$ is fixed
(due to FA events), the
MSEs of both data-aided sensing and 
repeated random sensing decrease with $K$
as shown in Fig.~\ref{Fig:plt4} (a).
It is also shown that 
the performance gap between
data-aided sensing and repeated random sensing, 
increases quickly when $K$ increases,
which demonstrates that 
data-aided sensing can perform better
than repeated random sensing
when there are more nodes thanks to the node selection criterion.

\begin{figure}[thb]
\begin{center}
\includegraphics[width=\figwidth]{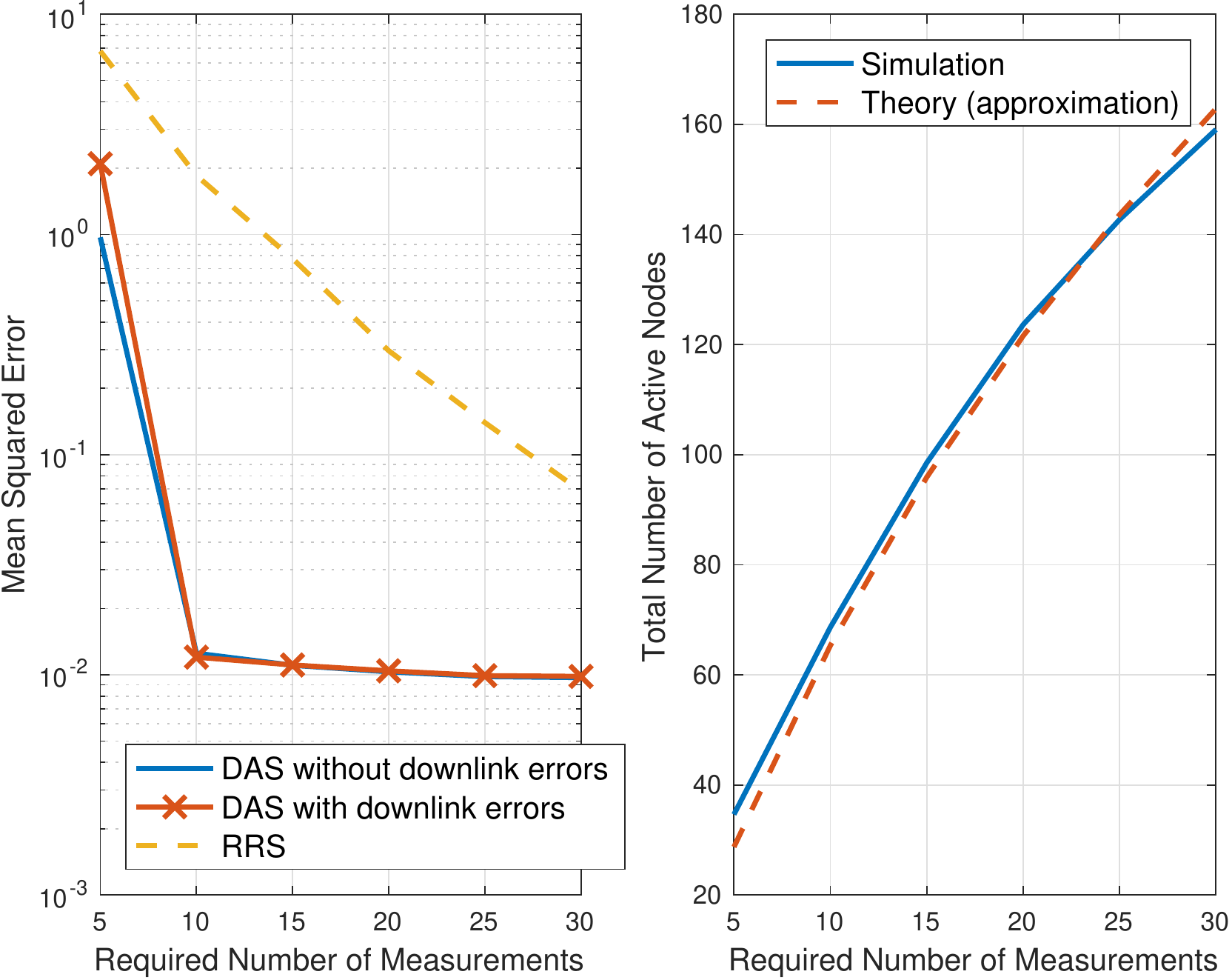} \\
\hskip 0.5cm (a) \hskip 3.5cm (b)
\end{center}
\caption{Performances of data-aided sensing and 
repeated random sensing for different numbers of 
$N$ when 
$K = 300$, $\gamma u_q = 0.5$ (for all $q$), 
and $S = 10$ (with a fixed number of rounds, $q = 3$):
(a) mean squared error, $\uE[||\bv - \bv_q||^2]$ with $q = 3$;
(b) the average total number of active nodes,
$\uE[\tilde N_0 + \ldots + \tilde N_q]$.}
        \label{Fig:plt6}
\end{figure}

In general, a better performance is expected with more measurements
in both data-aided sensing and repeated random sensing. 
As shown in Fig.~\ref{Fig:plt6} (a),
the MSEs decrease with $N$, where
$K = 300$, $\gamma u_q = 0.5$ (for all $q$), 
and $S = 10$ (with a fixed number of rounds, $q = 3$).
As mentioned earlier, the actual number of 
active nodes, $\tilde N_q$, differs from 
$N_q = N$, while the increase of $N$
results in the increase of $\tilde N_q$,
which is shown in Fig.~\ref{Fig:plt6} (b). 
Thus, for successful CRA, $N$ cannot be large.
It is noteworthy that according to Fig.~\ref{Fig:plt6} (a),
a large $N$ is not required in data-aided sensing, since
the MSE becomes low 
once $N$ is sufficiently large (say $N = 10$, which results in
$\uE[\tilde N_q] \approx \frac{70}{4} = 17.5$, which is much
smaller than $L - 1 = 63$ according to Fig.~\ref{Fig:plt6} (b)).
In other words, with a smaller number of measurements,
data-aided sensing can perform better than
repeated random sensing thanks to 
the node selection criterion.

\section {Concluding Remarks}	\label{S:Conc}

In this paper, data-aided sensing
was studied as a cross-layer approach to collect
data (or measurements) from IoT devices based
on the notion of CS when measurements
have a sparse representation.
A node selection criterion was proposed
to choose certain nodes in iterations for data-aided sensing
and it was shown that CRA plays a crucial role 
in collecting measurements efficiently.
Erroneous decisions at nodes 
and their impact on the performance have also been investigated
when the AP sends compressive transmission request.
Simulation results have shown that 
data-aided sensing outperforms repeated random sensing
thanks to active node selection through iterations.
In addition, it was demonstrated that although there are
downlink decision errors at nodes on transmission request,
their impact on the performance can be negligible
as the number of nodes increases and/or the sparsity
of the target signal is sufficiently low.

In this paper, data-aided sensing was applied to 
signals  that have sparse representations. As a further work,
a generalization of data-aided sensing to 
signals that do not have sparse representations
might be studied. It is also interesting to extend 
the notion of data-aided sensing 
to the upper layers and generalize it.
Furthermore, distributed data-aided sensing 
would be an important issue to support a large of devices/sensors.

\bibliographystyle{ieeetr}
\bibliography{sensor}

\end{document}